\documentstyle[prl,aps,epsf]{revtex}  
\begin{document}                
%
%
%
%
%
%

\newcommand{\CDF}{{\sc cdf}}
\newcommand{\FNAL}{{\sc fnal}}
\newcommand{\CLEO}{{\sc cleo}}
\newcommand{\SLAC}{{\sc slac}}
\newcommand{\LEP}{{\sc lep}}
\newcommand{\BABAR}{$\cal B${\em a}$\overline{\cal B}${\em ar}}
\newcommand{\DZero}{${\rm D}\emptyset$}
\newcommand{\CERN}{{\sc cern}}
\newcommand{\Belle}{{\sc Belle}}
\newcommand{\UAone}{{\sc ua1}}
\newcommand{\OPAL}{{\sc Opal}}
\newcommand{\HeraB}{{\sc Hera}-$B$}

\newcommand{\RunI}{Run~I}
\newcommand{\RunIa}{Run~I\,A}
\newcommand{\RunIb}{Run~I\,B}
\newcommand{\RunII}{Run~II}

\newcommand{\CDFNote}[1]{{\sc cdf}~Note~\##1}

\newcommand{\CTC}{{\sc ctc}}
\newcommand{\SVX}{{\sc svx}}
\newcommand{\SVXp}{{\sc svx}$^\prime$}
\newcommand{\VTX}{{\sc vtx}}
\newcommand{\CDT}{{\sc cdt}}
\newcommand{\CPR}{{\sc cpr}}

\newcommand{\CFT}{{\sc cft}}

\newcommand{\CMU}{{\sc cmu}}
\newcommand{\CMP}{{\sc cmp}}
\newcommand{\CMX}{{\sc cmx}}
\newcommand{\CSX}{{\sc csx}}
\newcommand{\CMUX}{{\sc cmu/x}}
\newcommand{\FMU}{{\sc fmu}}

\newcommand{\CEM}{{\sc cem}}
\newcommand{\CHA}{{\sc cha}}
\newcommand{\WHA}{{\sc wha}}
\newcommand{\FEM}{{\sc fem}}
\newcommand{\FHA}{{\sc fha}}
\newcommand{\PEM}{{\sc pem}}
\newcommand{\PHA}{{\sc pha}}

\newcommand{\BBC}{{\sc bbc}}

\newcommand{\TDC}{{\sc tdc}}
\newcommand{\MX}{{\sc mx}}
\newcommand{\FRC}{{\sc frc}}

\newcommand{\HTDC}{{\sc htdc}}
\newcommand{\CTCX}{{\sc ctcx}}

\newcommand{\CMUO}{{\sc cmuo}}
\newcommand{\TRKS}{{\sc trks}}
\newcommand{\TCMD}{{\sc tcmd}}
\newcommand{\TCMQ}{{\sc tcmq}}
\newcommand{\VTVZ}{{\sc vtvz}}

\newcommand{\CTVMFT}{{\sc ctvmft}}
\newcommand{\DIMUTG}{{\sc dimutg}}
\newcommand{\YBOS}{{\sc ybos}}

\newcommand{\BAnd}{{\sc .and.}}
\newcommand{\BOr}{{\sc .or.}}

\newcommand{\LI}{Level~1}
\newcommand{\LII}{Level~2}
\newcommand{\LIII}{Level~3}

%
%
\newcommand{\idest}{{\em i.e.}}

%
%

\newcommand{\KSht}{$K_{\rm S}$}

\newcommand{\BToVV}{$B \rightarrow V\,V$}

\newcommand{\BdDec}{$B^0 \rightarrow J/\psi\,K^{*0}$}
\newcommand{\BsDec}{$B_{\rm s}^0 \rightarrow J/\psi\,\phi$}

\newcommand{\BdGoldCP}{$B \rightarrow J/\psi\,K_{\rm S}$}

\newcommand{\BdKsDec}{$B^0 \rightarrow J/\psi\,K_{\rm S}$}
\newcommand{\BdKpDec}{$B^+ \rightarrow J/\psi\,K^+$}

\newcommand{\KsDec}{$K_{\rm S} \rightarrow \pi^+\,\pi^-$}
\newcommand{\KSCPDec}{$K^{*0} \rightarrow K_{\rm S}\,\pi^0$}
\newcommand{\KStarDec}{$K^{*} \rightarrow K\,\pi$}
\newcommand{\PhiDec}{$\phi \rightarrow K\,K$}

\newcommand{\BdToKPi}{$B^0 \rightarrow J/\psi\,K^+\,\pi^-$}

\newcommand{\pbarp}{$\overline{p}p$}

\newcommand{\RatBR}{${\cal B}(B^0 \rightarrow J/\psi\,K^{*0})\,/\,{\cal B}(B^+ \rightarrow J/\psi\,K^+)$}

%
%

\newcommand{\eeV}{${\rm eV}$}

\newcommand{\ekeV}{${\rm keV}$}
\newcommand{\pkeV}{${\rm keV}/c$}
\newcommand{\mkeV}{${\rm keV}/c^2$}

\newcommand{\eMeV}{${\rm MeV}$}
\newcommand{\pMeV}{${\rm MeV}/c$}
\newcommand{\mMeV}{${\rm MeV}/c^2$}

\newcommand{\eGeV}{${\rm GeV}$}
\newcommand{\pGeV}{${\rm GeV}/c$}
\newcommand{\mGeV}{${\rm GeV}/c^2$}

\newcommand{\eTeV}{${\rm TeV}$}
\newcommand{\pTeV}{${\rm TeV}/c$}
\newcommand{\mTeV}{${\rm TeV}/c^2$}

\newcommand{\mbarn}{${\rm mb}$}
\newcommand{\mubarn}{${\rm \mu b}$}

\newcommand{\tsec}{${\rm s}$}

\newcommand{\tpsec}{${\rm ps}$}
\newcommand{\xpsec}{${\rm ps}/c^2$}
\newcommand{\tnsec}{${\rm ns}$}
\newcommand{\xnsec}{${\rm ns}/c^2$}
\newcommand{\tmusec}{${\rm \mu s}$}
\newcommand{\tmsec}{${\rm ms}$}

\newcommand{\MHz}{${\rm MHz}$}
\newcommand{\kHz}{${\rm kHz}$}
\newcommand{\Hz}{${\rm Hz}$}

\newcommand{\micron}{${\rm \mu m}$}
\newcommand{\mm}{${\rm mm}$}
\newcommand{\cm}{${\rm cm}$}
\newcommand{\m}{${\rm m}$}
\newcommand{\Tesla}{${\rm T}$}
\newcommand{\degC}{$^\circ\,{\rm C}$}

\newcommand{\invpb}{${\rm pb}^{-1}$}

\newcommand{\pT}{$p_{\rm T}$}

%
%
\newcommand{\GToGf}{\Gamma_\perp/\Gamma}
\newcommand{\GLoGf}{\Gamma_{\rm L}/\Gamma}
\newcommand{\GToGt}{\frac{\Gamma_\perp}{\Gamma}}
\newcommand{\GLoGt}{\frac{\Gamma_{\rm L}}{\Gamma}}

%
%

\newcommand{\nUnits}[2]{$#1$\,{#2}}
\newcommand{\nError}[2]{$#1\pm #2$}
\newcommand{\nUnErr}[3]{$(#1\pm #2)$\,{#3}}

%
%



%
%
\title{
\begin{flushright}
CDF/PHYS/BOTTOM/PUBLIC/5281 \\
FERMILAB-PUB-00/165-E
\end{flushright}
Measurement of the Decay Amplitudes of \\
       \BdDec\ and \BsDec\ Decays
} 
\author{\font\eightit=cmti8
\def\r#1{\ignorespaces $^{#1}$}
\hfilneg
\begin{sloppypar}
\noindent
T.~Affolder,\r {21} H.~Akimoto,\r {43}
A.~Akopian,\r {36} M.~G.~Albrow,\r {10} P.~Amaral,\r 7 S.~R.~Amendolia,\r {32} 
D.~Amidei,\r {24} K.~Anikeev,\r {22} J.~Antos,\r 1 
G.~Apollinari,\r {10}
% ...
\end{sloppypar}
\vskip .026in
\begin{center}
(CDF Collaboration)
\end{center}

\vskip .026in
\begin{center}
\r 1  {\eightit Institute of Physics, Academia Sinica, Taipei, Taiwan 11529, 
Republic of China} \\
\r 2  {\eightit Argonne National Laboratory, Argonne, Illinois 60439} \\
\r 3  {\eightit Istituto Nazionale di Fisica Nucleare, University of Bologna,
I-40127 Bologna, Italy} \\
\r 4  {\eightit Brandeis University, Waltham, Massachusetts 02254} \\
\r 5  {\eightit University of California at Los Angeles, Los 
Angeles, California  90024} \\  
\r 6  {\eightit Instituto de Fisica de Cantabria, CSIC-University of Cantabria, 
39005 Santander, Spain} \\
\r 7  {\eightit Enrico Fermi Institute, University of Chicago, Chicago, 
Illinois 60637} \\
\r 8  {\eightit Joint Institute for Nuclear Research, RU-141980 Dubna, Russia}
\\
\r 9  {\eightit Duke University, Durham, North Carolina  27708} \\
\r {10}  {\eightit Fermi National Accelerator Laboratory, Batavia, Illinois 
60510} \\
\r {11} {\eightit University of Florida, Gainesville, Florida  32611} \\
\r {12} {\eightit Laboratori Nazionali di Frascati, Istituto Nazionale di Fisica
               Nucleare, I-00044 Frascati, Italy} \\
\r {13} {\eightit University of Geneva, CH-1211 Geneva 4, Switzerland} \\
\r {14} {\eightit Harvard University, Cambridge, Massachusetts 02138} \\
\r {15} {\eightit Hiroshima University, Higashi-Hiroshima 724, Japan} \\
\r {16} {\eightit University of Illinois, Urbana, Illinois 61801} \\
\r {17} {\eightit The Johns Hopkins University, Baltimore, Maryland 21218} \\
\r {18} {\eightit Institut f\"{u}r Experimentelle Kernphysik, 
Universit\"{a}t Karlsruhe, 76128 Karlsruhe, Germany} \\
\r {19} {\eightit Korean Hadron Collider Laboratory: Kyungpook National
University, Taegu 702-701; Seoul National University, Seoul 151-742; and
SungKyunKwan University, Suwon 440-746; Korea} \\
\r {20} {\eightit High Energy Accelerator Research Organization (KEK), Tsukuba, 
Ibaraki 305, Japan} \\
\r {21} {\eightit Ernest Orlando Lawrence Berkeley National Laboratory, 
Berkeley, California 94720} \\
\r {22} {\eightit Massachusetts Institute of Technology, Cambridge,
Massachusetts  02139} \\   
\r {23} {\eightit Institute of Particle Physics: McGill University, Montreal 
H3A 2T8; and University of Toronto, Toronto M5S 1A7; Canada} \\
\r {24} {\eightit University of Michigan, Ann Arbor, Michigan 48109} \\
\r {25} {\eightit Michigan State University, East Lansing, Michigan  48824} \\
\r {26} {\eightit University of New Mexico, Albuquerque, New Mexico 87131} \\
\r {27} {\eightit The Ohio State University, Columbus, Ohio  43210} \\
\r {28} {\eightit Osaka City University, Osaka 588, Japan} \\
\r {29} {\eightit University of Oxford, Oxford OX1 3RH, United Kingdom} \\
\r {30} {\eightit Universita di Padova, Istituto Nazionale di Fisica 
          Nucleare, Sezione di Padova, I-35131 Padova, Italy} \\
\r {31} {\eightit University of Pennsylvania, Philadelphia, 
        Pennsylvania 19104} \\   
\r {32} {\eightit Istituto Nazionale di Fisica Nucleare, University and Scuola
               Normale Superiore of Pisa, I-56100 Pisa, Italy} \\
\r {33} {\eightit University of Pittsburgh, Pittsburgh, Pennsylvania 15260} \\
\r {34} {\eightit Purdue University, West Lafayette, Indiana 47907} \\
\r {35} {\eightit University of Rochester, Rochester, New York 14627} \\
\r {36} {\eightit Rockefeller University, New York, New York 10021} \\
\r {37} {\eightit Rutgers University, Piscataway, New Jersey 08855} \\
\r {38} {\eightit Texas A\&M University, College Station, Texas 77843} \\
\r {39} {\eightit Texas Tech University, Lubbock, Texas 79409} \\
\r {40} {\eightit Istituto Nazionale di Fisica Nucleare, University of Trieste/
Udine, Italy} \\
\r {41} {\eightit University of Tsukuba, Tsukuba, Ibaraki 305, Japan} \\
\r {42} {\eightit Tufts University, Medford, Massachusetts 02155} \\
\r {43} {\eightit Waseda University, Tokyo 169, Japan} \\
\r {44} {\eightit University of Wisconsin, Madison, Wisconsin 53706} \\
\r {45} {\eightit Yale University, New Haven, Connecticut 06520} \\
\end{center}

 }
\address{Submitted to PRL, 12 July 2000}
\maketitle
\begin{abstract}                

A full angular analysis has been performed for the 
pseudo-scalar to vector-vector decays \BdDec\ and \BsDec\
to determine
the amplitudes for decays with parity--even longitudinal ($A_0$) and 
transverse ($A_\parallel$) polarization 
and parity--odd
transverse ($A_\perp$) polarization.
The measurements are based on 190 $B^0$~candidates and 40
$B_{\rm s}^0$~candidates collected from a data set corresponding to
$89\,{\rm pb}^{-1}$ of $\bar{\rm p}$p collisions at $\sqrt s = 1.8$ TeV
at the Fermilab Tevatron.
In both decays the decay amplitude for longitudinal polarization
dominates : 
$|A_0|^2 = 0.59 \pm 0.06 \pm 0.01$
for the $B^0$ decay and 
$|A_0|^2 = 0.61 \pm 0.14 \pm 0.02$
for the $B_{\rm s}^0$ decay.  The parity--odd amplitude  is found to
be small: 
$|A_\perp|^2 = 0.13 {+0.12 \atop -0.09} \pm 0.06$ 
for $B^0$ and 
$|A_\perp|^2 = 0.23 \pm 0.19 \pm 0.04$ 
for $B_{\rm s}^0$.

\vspace{0.1in}

\end{abstract}

\bigskip


The decays \BdDec\ and \BsDec\
are pseudo-scalar to vector-vector decays and in
principle have three decay amplitudes which can be determined by
studying the angular distributions of the final state particles.
These decays can have orbital angular momenta between the
$J/\psi$ and $K^{\ast} $ (or $\phi$) of 0, 1, or 2,
and three decay amplitudes govern these transitions.
Another convenient description 
is given in the transversity basis~\cite{Lipkin,DigDunLipRos} in which
the decay amplitudes are defined in terms
of the linear polarization of the vector mesons.
In this basis the $L = 1$  (P wave) decays are governed by a single
decay amplitude, $A_\perp$, corresponding to a parity--odd correlation
between transversely polarized vector mesons. The other two decay amplitudes,
$A_0$
and $A_\parallel$, are combinations of the parity--even $L = 0$ and $L = 2$
(S and D wave) decays. The longitudinal
polarization fraction, as is commonly defined
in the helicity
basis~\cite{VV}, is given by $\Gamma_L /\Gamma = |A_0|^2$.

Determination of the decay amplitudes and phases
probes the limitations of 
the factorization hypothesis~\cite{factorization}.  
Factorization
assumes that a weak decay matrix element can be described as
the product of two independent (hadronic) currents,
in this case 
treating the $J/\psi$ and $B
\rightarrow K^{\ast}$ ($\phi$) as currents. To the extent that
factorization holds,
final state interactions are negligible.  The matrix elements
factorize into short and long distance (weak and strong) parts
and the polarization decay amplitudes do
not interfere and so are expected to be relatively real.

A measurement of the parity--odd amplitude, $A_\perp$, is pertinent
to studies of $CP$ invariance.
For example, if the decay $B^0 \rightarrow J/\psi \, K^{\ast 0}$
(with $K^{\ast 0} \rightarrow
K_{\rm s}^0 \, \pi^0$) 
were to occur mainly through either a parity--odd or even amplitude,
then this mode
could be used~\cite{Lipkin,Kayser} as readily as
the $B^0 \rightarrow J/\psi \, K_{\rm s}^0$ decay mode
for determining\cite{multitag} the $CP$ nonconserving parameter $\sin (2 \beta )$.
The situation holds
as well for the decay $B_{\rm s}^0 \rightarrow J/\psi \phi$,
which is expected to have a very small $CP$ decay rate asymmetry in
the Standard (CKM) Model.
In addition
the polarization of the decay \BsDec\ is interesting for
its potential to improve the precision of measurements
of the lifetime difference between the $B_{\rm s}^0$ mass eigenstates via 
an angular analysis~\cite{DigDunLipRos,deltagamma}.

The longitudinal polarization in \BdDec\ was first measured by
ARGUS~\cite{HAlb_plb_340_217} as $\sim$1.0
followed by a
CLEO~\cite{MSAlam_prd_50_43} measurement
of $0.8 \pm 0.2$ and a CDF (Collider Detector at Fermilab)
measurement~\cite{FAbe_prl_75_3069} of
\nError{0.65}{0.11}. A full angular analysis by
CLEO~\cite{CPJess_prl_79_4533} obtained a longitudinal polarization
fraction of \nError{0.52}{0.08} and a parity--odd
fraction of \nError{0.16}{0.09}.
The only  previous measurement of the $B_{\rm s}^0$ decay mode, by
CDF~\cite{FAbe_prl_75_3069}, obtained a longitudinal polarization
fraction of \nError{0.56}{0.21}, consistent with 
the results for $B^0$ as expected under $SU(3)$-flavor
symmetry. 

This Letter describes a full angular analysis of both the $B^0$
and $B_{\rm s}^0$ decay modes based on $89\,{\rm pb}^{-1}$ of
$\bar{p}p$ collisions at
$\sqrt{s} =$\nUnits{1.8}{\eTeV} collected with the CDF detector
at the Fermilab Tevatron.
In this analysis, the \BdDec\ and \BsDec\ decays are reconstructed
from the decay modes $J/\psi \rightarrow \mu^+ \mu^-$, $K^{*0} \rightarrow
K^+ \pi^-$, and $\phi \rightarrow K^+ K^-$.
After the selection described below, there are 190 \BdDec\ and 40
\BsDec\ candidates above background. The results presented here are
independent of those in Ref.~\cite{FAbe_prl_75_3069}.

CDF is a general purpose detector 
and has been described in detail elsewhere~\cite{BlueBook}.
For this analysis important elements of the detector
are the silicon vertex detector, with track
impact parameter resolution of $10$\micron, and the drift
chamber, with charged particle momentum resolution
of $\delta p_T/p_T\,\sim\,0.001\,p_T$,
where \pT ~(in \pGeV) is the component of the momentum
transverse to the $\bar{p}p$ collision axis.  Chambers for
muon detection provide coverage for particles 
with direction within 40 degrees of
the transverse direction. 

The event sample for this measurement is selected
by online trigger and offline reconstruction criteria.
The
trigger signature is the decay of a $J/\psi$
to two muons. 
Hits in the muon chambers forming a track segment consistent with a
nominal muon transverse momentum 
\pT$>$\nUnits{3.0}{\pGeV}
satisfy the first level of the trigger. Candidate muon
track segments must be separated
by more than 10$^\circ$ in the plane transverse to the collision axis.
The second level requires drift chamber tracks, with
\pT$>$\nUnits{2}{\pGeV}, which extrapolate to
the track segments in the muon chambers.  
The third level accepts $J/\psi$ candidates with a reconstructed mass
between \nUnits{2.8} and \nUnits{3.4}{\mGeV}. 

The offline reconstruction first selects $J/\psi$ candidates.
Two muons of opposite charge satisfying quality
requirements are combined into a $J/\psi$ candidate. A kinematic
fit requiring the muons to originate from a common 
vertex is performed, and the confidence level of the
fit is required to be at least 0.01.
This results in a sample of about 290,000
candidate $J/\psi$'s.

Of these, candidates within \nUnits{80}{\mMeV} of the 
 $J/\psi$ mass (\nUnits{3096.88}{\mMeV})~\cite{PDG}
are then combined with two oppositely charged
tracks that form a $K^*$ ($\phi$)
for the \BdDec\ (\BsDec\ ) decays.
CDF lacks the particle identification  capability to distinguish
between the two
$K$-$\pi$ charge assignments for a $K^*$ candidate.
However, the mass distribution for misidentified
candidates is broader than the natural width of the $K^*$
and their contribution
is largely suppressed by choosing
the assignment yielding a
$K^*$ mass closer to,
and within {\nUnits{80}{\mMeV}
of, the
world average (\nUnits{896.1}{\mMeV}).

To further improve the signal to noise ratio, the $B^0$ candidate is
required to have \pT$ > $\nUnits{6.0}{\pGeV}, with the $K^*$ having
\pT$ > $\nUnits{2.0}{\pGeV}.
All four particles from the $B^0$ decay are required to come from
a common `secondary' vertex, with the confidence level of the fit being
greater than 0.001.
The proper decay length ($c \times$ proper decay time) for the
$B^0$ is required to be at least
\nUnits{100}{\micron}.
The resulting $K \pi \mu^+ \mu^-$ mass
distribution is plotted in Figure~\ref{fig:mplots}(a).

For the $B_{\rm s}^0$ decay, the $\phi$
candidate is required to have a
mass within \nUnits{10}{\mMeV} of the nominal $\phi$
mass (\nUnits{1019.4}{\mMeV}). The 
narrow natural width of the $\phi$ provides for 
better background rejection than
the $K^*$. In order to compensate for
the lower expected yield of $B^0_s$ events, looser requirements are
imposed:
transverse momentum cuts of \nUnits{1.5}{\pGeV} and
\nUnits{4.5}{\pGeV} are applied to the $\phi$ and $B_{\rm s}^0$ candidates.
The confidence level of the four track fit is again required
to be greater than 0.001, and the proper decay length
is required to be at least
\nUnits{50}{\micron}.
The resulting $K K \mu^+ \mu^-$ mass
distribution
is plotted in Figure~\ref{fig:mplots}(b).

The linear polarization of the vector mesons is determined from the
angular distribution of the final state decay products.
The decay angles are defined~\cite{DigDunLipRos} as $\Theta_{K^*}$, 
$\Theta_{\rm T}$, and $\Phi_{\rm T}$.
In the rest frame of the $K^*$,
$\Theta_{K^*}$ 
is the angle between the $K$ momentum and the direction opposite
the $B$ meson (for the $\phi$, the $K^+$ defines the angle).
The transversity angles, $\Theta_{\rm T}$ and $\Phi_{\rm T}$, are
defined in the $J/\psi$ rest frame: $\Theta_{\rm T}$ is the angle
between the $\mu^+$ momentum and the perpendicular to the $K^*$-$K$
plane; $\Phi_{\rm T}$ is the azimuthal angle from the $K^*$ to the
projection of the $\mu^+$ momentum onto the $K^*$-$K$ plane.

The decay angular distribution is~\cite{deltagamma}:

\begin{displaymath}
 \begin{array}[c]{c}
   d\Gamma/d\Omega \propto
   \begin{array}[t]{cl}
       & 2 \cos^2 \Theta_{K^*} ( 1 - \sin^2 \Theta_{\rm T} \cos^2
         \Phi_{\rm T} ) \,|A_0|^2
 \\ + &
         \sin^2 \Theta_{K^*} ( 1 - \sin^2 \Theta_{\rm T} \sin^2
         \Phi_{\rm T} ) \,|A_\parallel|^2
 \\ + & \sin^2 \Theta_{K^*} \sin^2 \Theta_{\rm T}\,|A_\perp|^2
 \\ + & \frac{1}{\sqrt{2}}\sin 2\Theta_{K^*}\sin^2 \Theta_{\rm T}
         \sin 2\Phi_{\rm T}\,{\rm Re}\,(A_0^*A_\parallel)
 \\ \mp &
         \sin^2 \Theta_{K^*} \sin 2\Theta_{\rm T} \sin \Phi_{\rm
         T}\,{\rm Im}\,(A_\parallel^*A_\perp)
 \\ \pm &
         \frac{1}{\sqrt{2}}\sin 2\Theta_{K^*}\sin 2\Theta_{\rm T}
         \cos \Phi_{\rm T}\,{\rm Im}\,(A_0^*A_\perp)
   \end{array}
 \end{array}
 \label{equ:ang_dist}
\end{displaymath}

\noindent Normalization of the distribution to unity implies
$|A_0|^2 + |A_\parallel|^2 + |A_\perp|^2 = 1$, and an
unobservable phase is removed by requiring $\arg(A_0) = 0$.
This leaves four measurable quantities: the
polarization fractions, $\GLoGf = |A_0|^2$ and $\GToGf = |A_\perp|^2$,
and the phases of the matrix elements, $\arg(A_\parallel)$ and
$\arg(A_\perp)$. 
The last
two terms of the angular distribution
have opposite signs for particle and anti-particle decay.
The $B^0$ and $\overline{B}^0$ decays are flavor tagged
by the charge of the $K$ meson. The $B_{\rm s}^0$ and
$\overline{B}_{\rm s}^0$ are not
distinguishable by their final state particles, so for $B_{\rm s}^0$
decays information about the phase of $A_\perp$ is lost.

The matrix elements are extracted by
fitting the observed decay
angular distribution.
The fit is performed over the decay
angles and the $B$ candidate mass range covered in Figure 1. 
To account for detector acceptance and selection criteria,
Monte Carlo events are
generated with detector and trigger simulations and processed
by the same reconstruction software used on the data.
The fit method uses an unbinned
log-likelihood
with the normalization depending on the parameters
in the fit~\cite{pappas_thesis,fitmethod}. 

The background is modeled by
a sum of polynomial terms of $\cos\,\Theta_{K^*}$ and
$\cos\,\Theta_{\rm T}$, and sines and cosines of $\Phi_{\rm T}$ and
$2\Phi_{\rm T}$.  
The background angular distribution is determined from the
events on both sides of the $B$ meson mass peak.

Additional terms are included in the angular distribution to account
for the residual probability ($\sim$6\%) of misidentifying the final state hadrons
in the selected \BdDec\ decays.
Events reconstructed under the wrong hypothesis will
yield incorrect decay angles having a distribution different from, but
fully correlated with, the polarized decay distribution.
The augmented likelihood function corrects for the effect based on
the kinematics as determined from Monte Carlo. 

The results from the global fit are illustrated in Figure~\ref{fig:projs}. 
The independent variables are functions of the transversity angles:
$\cos\,\Theta_{K^*}$, $\cos\,\Theta_{\rm T}$, and $\Phi_{\rm T}$.
The points
are the background subtracted projection of the data corrected for detector and
reconstruction efficiencies.  The superimposed curves, derived from
the results of the global
fit, are in good agreement with the data.

Sources of systematic uncertainty are, in order of importance,
the model of the background shape,
Monte Carlo $B$ generation parameters, trigger simulation,
and $B^0$-$B_{\rm s}^0$ cross talk.
The last two are found to have negligible contribution.
The uncertainty of the final
result is still dominated by the statistics of the
event sample~\cite{pappas_thesis}.

The underlying background shape is not known from first
principles.
To estimate the systematic uncertainty from the background modelling,
multiple fits with different parametrizations of the background
decay angles were studied.
Different models of the shape are sensitive to different
inaccuracies of the background shape function, and
the spread in the fitted values
is used to estimate the sensitivity to possibly unaccounted 
structure.

The input parameters to the Monte Carlo generation of $B$ mesons 
are the $b$ quark mass and generation mass scale, the parton
distribution functions and fragmentation parameters. The
effect on the measurement is determined by varying the parameters by
their nominal uncertainties and refitting for the decay amplitudes.
Systematic uncertainties due to the modeling of the trigger selection
criteria are studied in a similar fashion.

The similarity of the decay
kinematics for the $B^0$ and $B_{\rm s}^0$ decays
allows for a possible cross contamination through $K$-$\pi$ misidentification. 
Monte Carlo studies give contamination
fractions of $2.1\%$ ($4.4\%$) in the $B^0$ ($B_{\rm s}^0$) sample.
The assigned systematic uncertainty 
is equal to the
difference in the $B^0$ and $B_{\rm s}^0$ polarization fraction,
multiplied by the
contamination fraction. 

The final results of this analysis are summarized in
Table~\ref{tab:BVV_res}, and are illustrated in
Figure~\ref{fig:result}. For the \BdDec\ the
magnitudes of decay amplitudes are in good agreement
with the results from CLEO~\cite{CPJess_prl_79_4533}.
The longitudinal polarization fraction dominates: 
$\GLoGf = |A_0|^2 = 0.59 \pm 0.06 \pm 0.01$
and the parity--odd fraction is small: 
$\GToGf = |A_\perp|^2 = 0.13 {+0.12 \atop -0.09} \pm 0.06$.
In contrast to the CLEO result, the CDF measurement
for the phase of $A_\parallel$
favors the presence of strong final state interactions,
in as much as $\arg(A_\parallel)$ is not close to $0$ or $\pi$.
However, the results from the two experiments are
not inconsistent, given their uncertainties. 
The results for the \BsDec\ are the first
available from a full angular analysis.  The $B_{\rm s}^0$ results
also show a dominant longitudinal polarization fraction
$\GLoGf = |A_0|^2 = 0.61 \pm 0.14 \pm 0.02$
and a small parity--odd fraction:
$\GToGf = |A_\perp|^2 = 0.23 \pm 0.19 \pm 0.04$,
consistent with the $B^0$ results and SU(3) flavor symmetry.

We thank the Fermilab staff and the technical staffs of
the participating institutions for their vital contributions.
This research was supported by the U. S. Department of
Energy and the National Science Foundation; the Italian Istituto
Nazionale di Fisica Nucleare; the Ministry of Education, Science and
Culture of Japan; the Natural Sciences and Engineering Research
Council of Canada; the National Science Council of the Republic of
China; the Swiss National Science Foundation; the A. P. Sloan Foundation;
and the Bundesministerium f\"{u}r Bildung und Forschung, Germany.
We thank H.~Lipkin for
bringing the transversity basis to our attention and acknowledge
illuminating correspondence and discussions with I.~Dunietz and J.~Rosner.

%


\begin{table}
   \begin{tabular}{lll}

	Quantity & \BdDec\ & \BsDec\ \\
\tableline

         $A_0$ & $0.77 \pm 0.04 \pm 0.01$ & $0.78 \pm 0.09 \pm 0.01$ \\


          $A_\parallel$  & $0.53 \pm 0.11 \pm 0.04$
			& $0.41 \pm 0.23 \pm 0.05$ \\

	$arg(A_\parallel)$ & $2.2 \pm 0.5 \pm 0.1$
			& $1.1 \pm 1.3 \pm 0.2$ \\


          $A_\perp$ & $0.36 \pm 0.16 \pm 0.08$
         		& $0.48 \pm 0.20 \pm 0.04$ \\

	$arg(A_\perp)$ & $-0.6 \pm 0.5 \pm 0.1$
			&  \\
\tableline


            $\GLoGf = |A_0|^2$ & $0.59 \pm 0.06 \pm 0.01$
			& $0.61 \pm 0.14 \pm 0.02$ \\
            

	$\GToGf = |A_\perp|^2$ &  $0.13 {+0.12 \atop -0.09} \pm 0.06$
			& $0.23 \pm 0.19 \pm 0.04$ 

   \end{tabular}
   \caption{Fitted decay amplitudes and phases (in radians).
           The uncertainties are
            statistical and systematic, respectively.}
   \label{tab:BVV_res}
\end{table}


\begin{figure}
   \centerline{
      \epsfxsize1.9in\epsffile{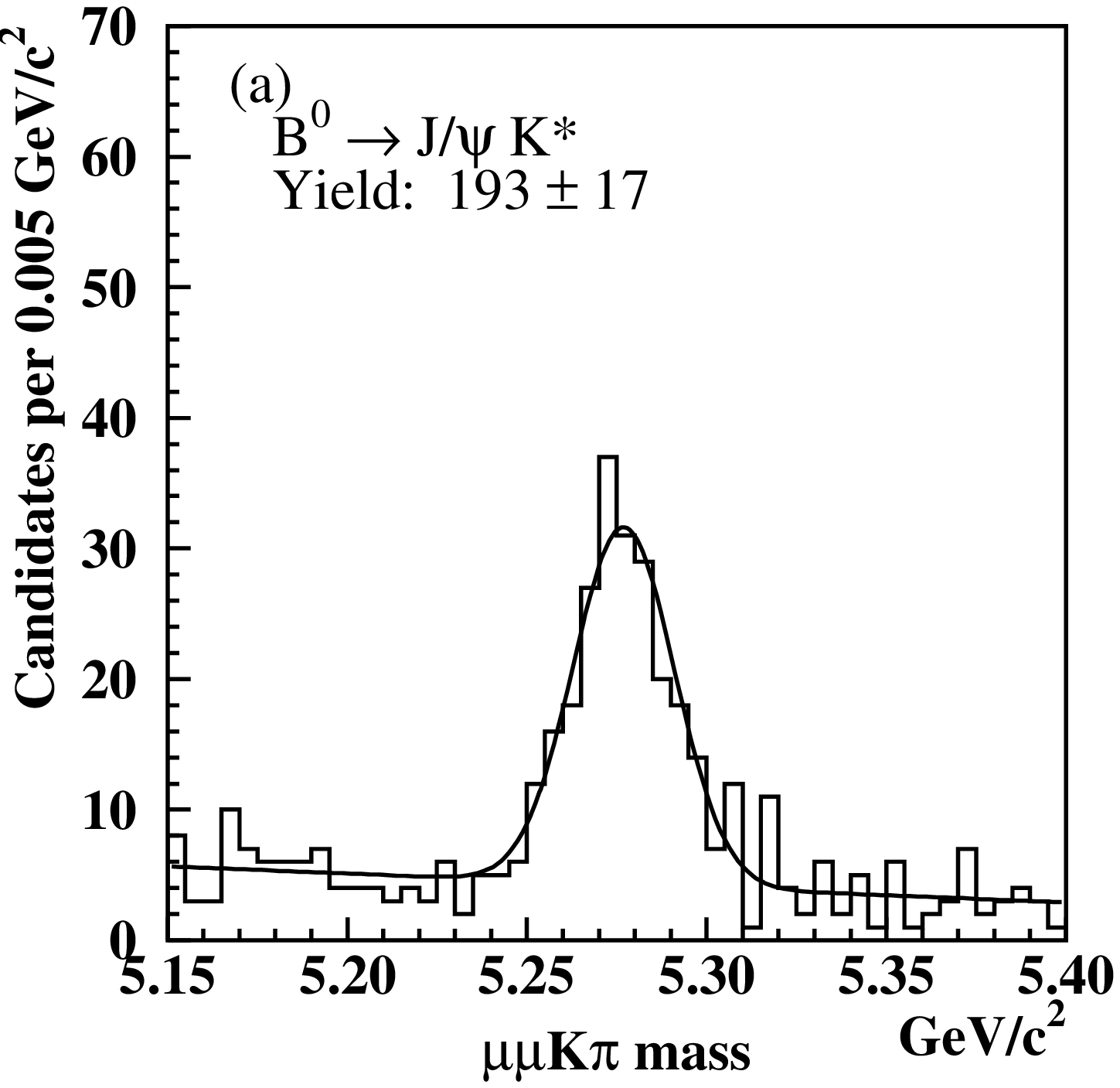}
	\hspace{-0.25in}
      \epsfxsize1.9in\epsffile{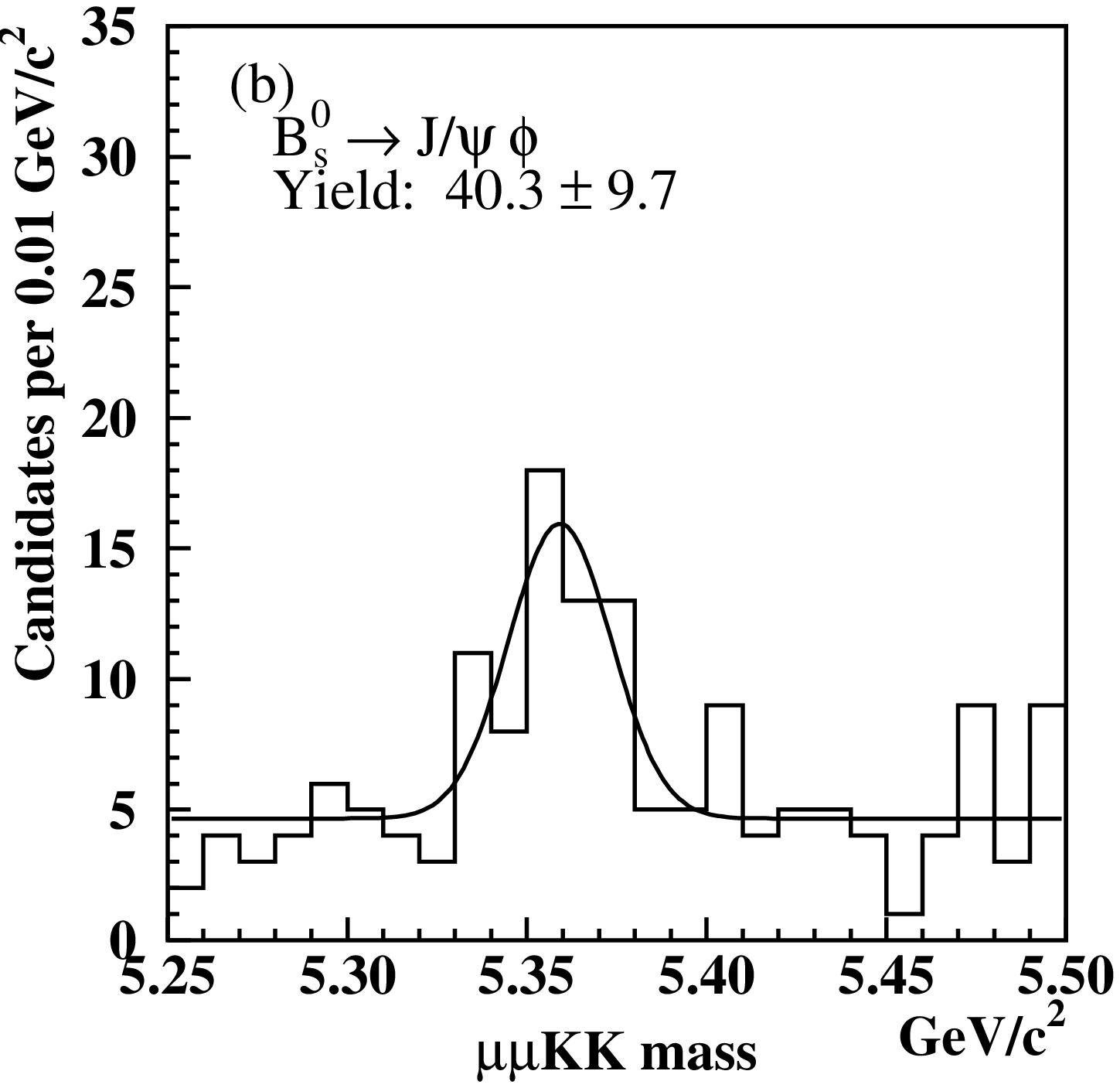}
   }
	\vspace{0.1in}
   \caption{Invariant mass distributions, after all selection requirements,
           for (a) \BdDec\  and
            (b) \BsDec\ candidates.}
   \label{fig:mplots}
\end{figure}


\begin{figure}
   \centerline{
               \epsfxsize3.6in\epsffile{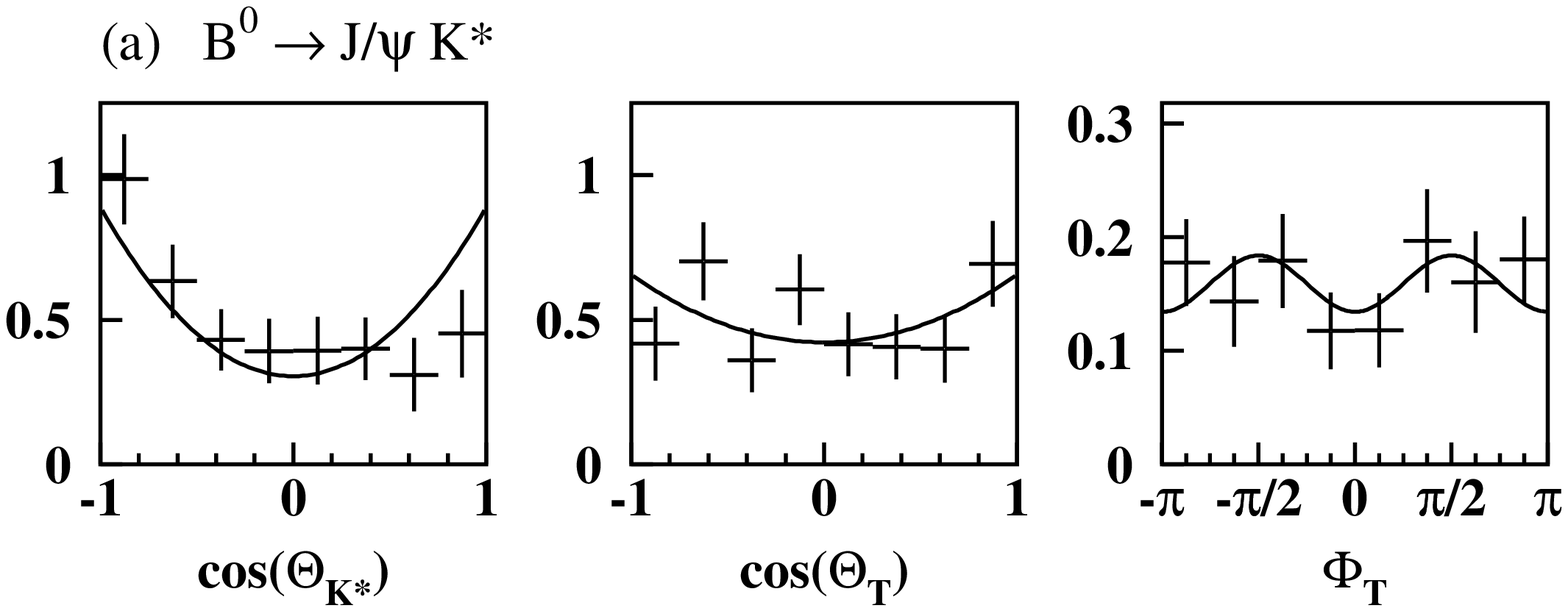}}
   \centerline{
               \epsfxsize3.6in\epsffile{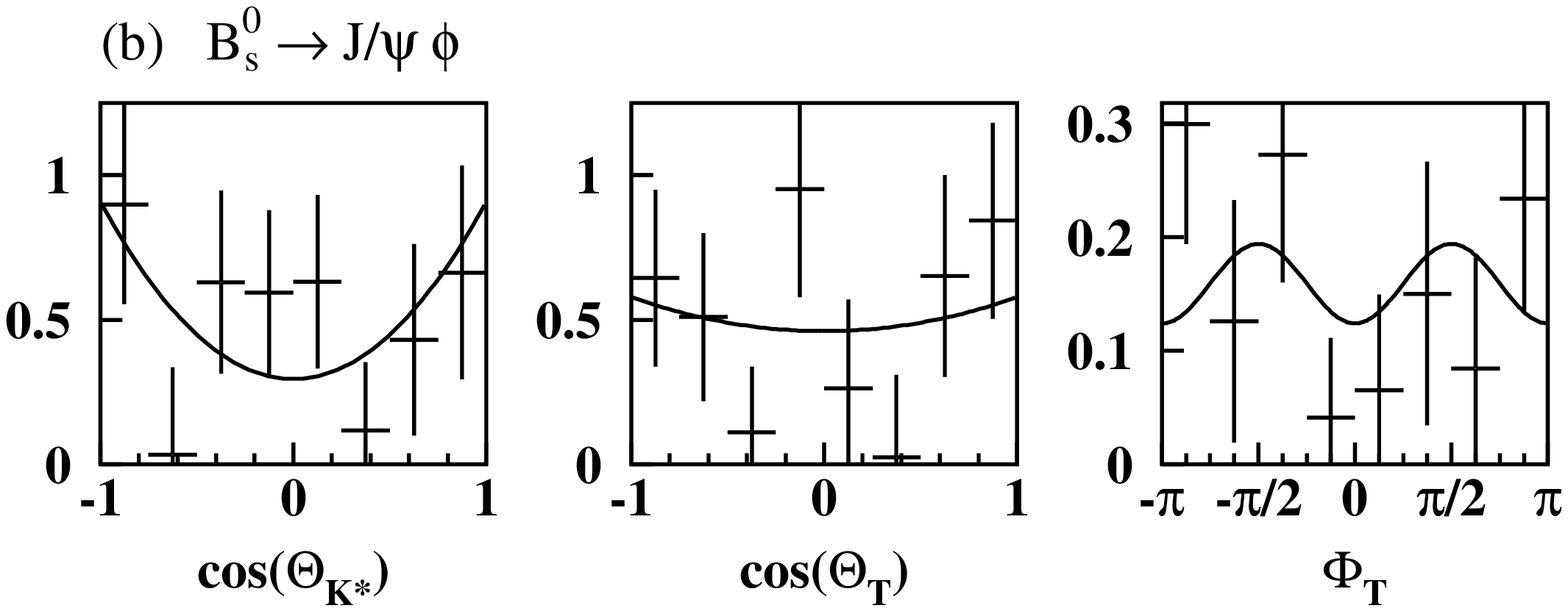}}
	\vspace{0.1in}
   \caption{Projections of the results from the full angular
           fit and the background subtracted
	    data, corrected for acceptance, are shown for each 
            of the decay angles, for (a) the \BdDec\ mode and (b)
		the \BsDec\ mode.}
   \label{fig:projs}
\end{figure}


\begin{figure}
   \centerline{\epsfxsize2.0in\epsffile{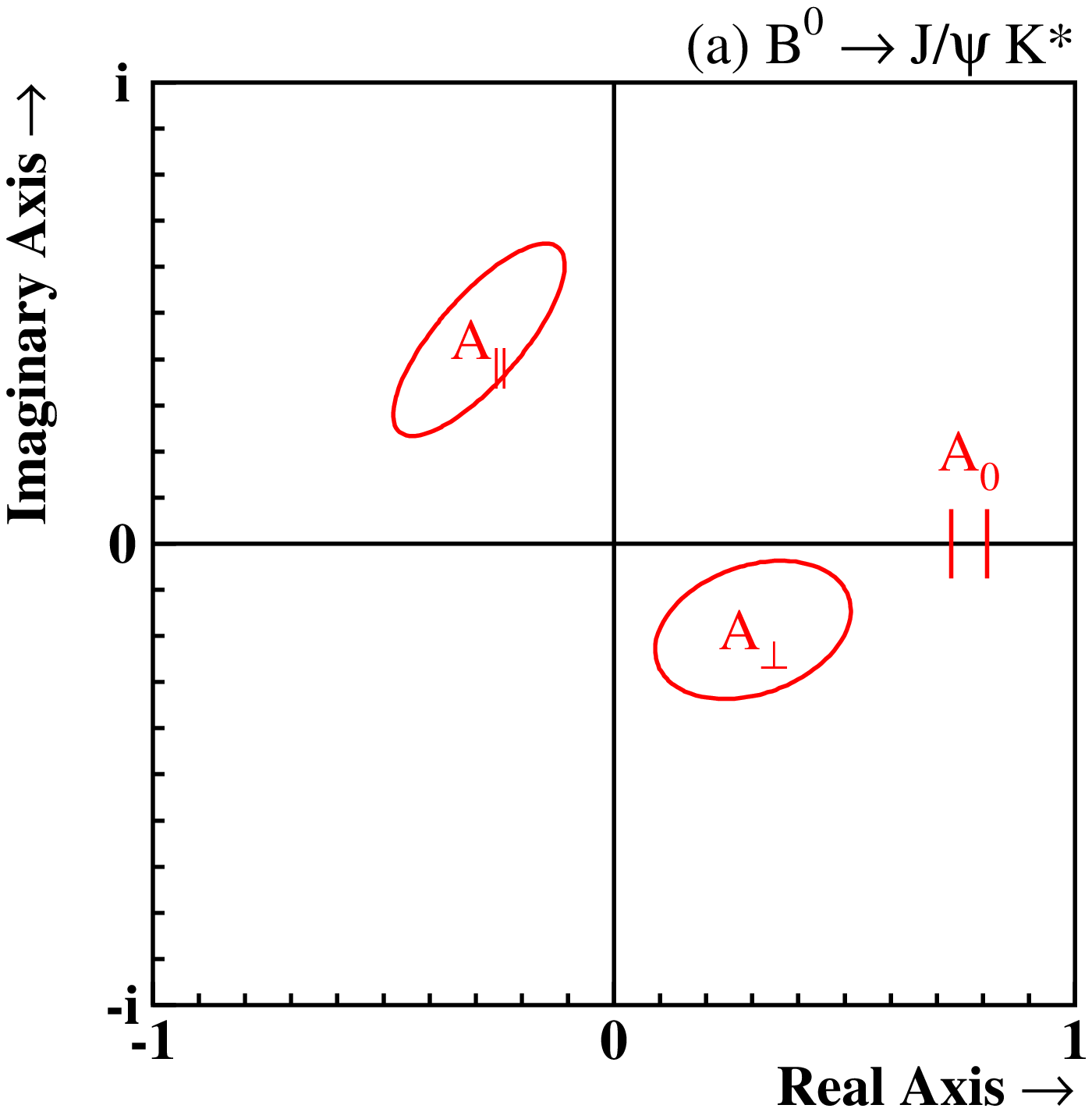}
	\hspace{-0.25in}
               \epsfxsize2.0in\epsffile{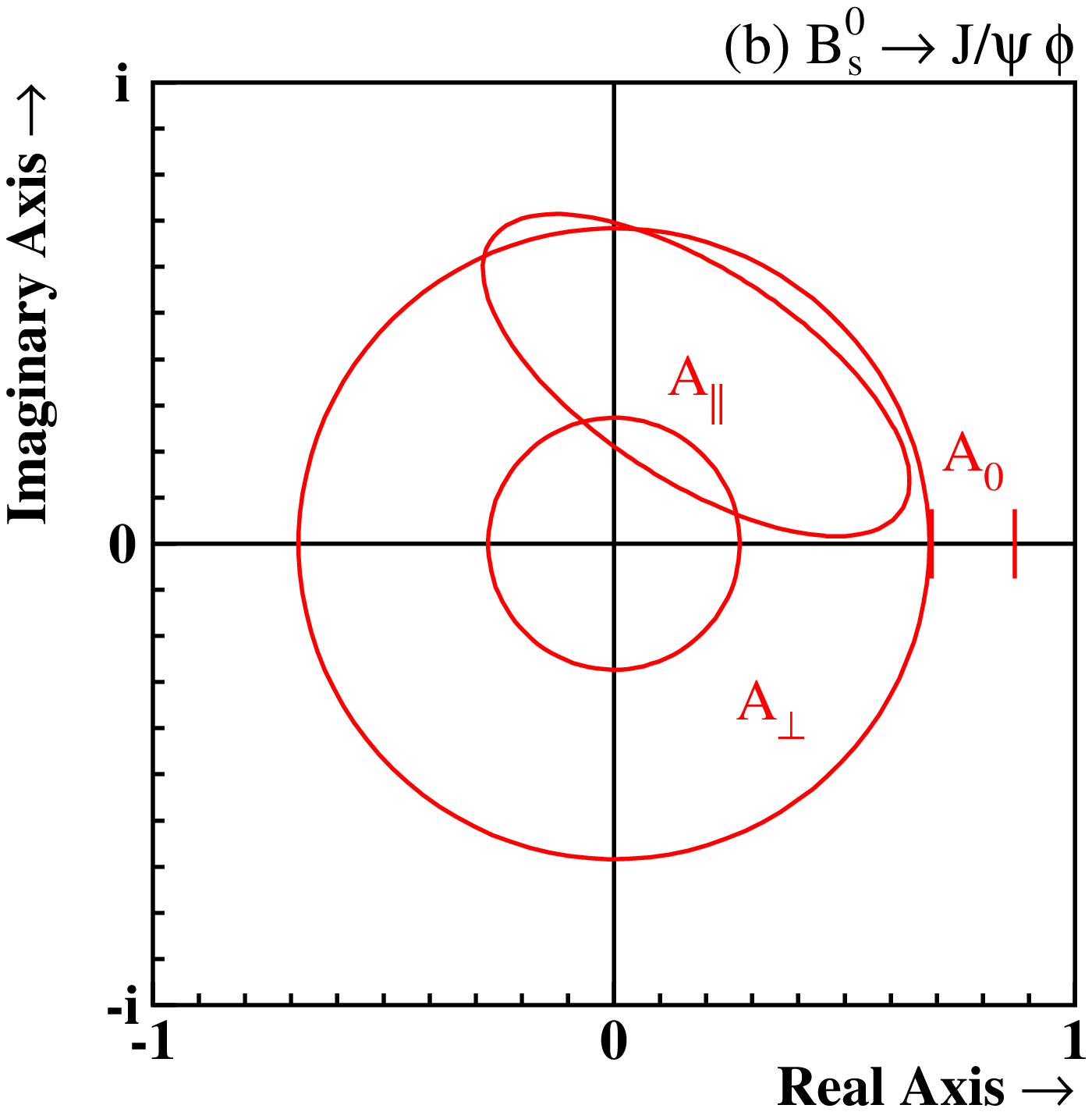}
   }
	\vspace{0.1in}
   \caption{One standard deviation
	contours, including statistical and systematic uncertainties,
	for the fitted decay amplitudes in the
            complex plane, for (a) \BdDec\ and
            (b) \BsDec\ decays. } 
   \label{fig:result}
\end{figure}

\end{document}